# Internal interface strains effects on $UO_2/U_3O_7$ oxidation behaviour

N. Creton[1,a], V. Optasanu[1,b], S. Garruchet[1,c], L. Raceanu[1,d],

T. Montesin[1,e], L. Desgranges[2,f], S. Dejardin[1,g]

[1]I.C.B., UMR 5209 CNRS, 9 Avenue Alain Savary, B.P. 47870 21078 Dijon, FRANCE

[2]CEA/DEN/DEC/SESC/LLCC, CEA Cadarache, 13115 St Paul lez Durance, FRANCE

[a]nicolas.creton@u-bourgogne.fr, [b]virgil.optasanu@u-bourgogne.fr,
[c]sebastien.garruchet@u-bourgogne.fr, [d]laura.raceanu@u-bourgogne.fr,
[e]tony.montesin@u-bourgogne.fr, [f]lionel.desgranges@cea.fr,
[g]steeve.dejardin@u-bourgogne.fr

**Abstract.** The growth of a $U_3O_7$ oxide layer during the anionic oxidation of $UO_2$ pellets induced very important mechanical stresses due to the crystallographic lattice parameters differences between $UO_2$ and its oxide. These stresses, combined with the chemical processes of oxidation, can lead to the cracking of the system, called chemical fragmentation. We study the crystallographic orientation of the oxide lattice growing at the surface of $UO_2$, pointing the fact that epitaxy relations at interface govern the coexistence of $UO_2$ and $U_3O_7$. In this work, several results are given:

- Determination of the epitaxy relations between the substrate and its oxide thanks to the Bollmann's method; epitaxy strains are deduced.
- Study of the coexistence of different domains in the $U_3O_7$ (crystallographic compatibility conditions at the interface between two phases: Hadamard conditions).
- FEM simulations of a multi-domain $U_3O_7$ connected to a $UO_2$ substrate explain the existence of a critical thickness of $U_3O_7$ layer.

**Introduction**

In order to have a better understanding of mechanisms that govern the chemical fragmentation reaction related to the anionic oxidation of $UO_2$ we propose here to study the relations between mechanical stress and oxygen diffusion in a system composed of $U_3O_7/UO_2$. This study need:
- to understand the nature of the interface existing between $UO_2$ and $U_3O_7$,
- to determine the interfacial stress-free strain tensor activated during the $UO_2$ to $U_3O_7$ transformation.

In a first part we study the oxidation front between these two phases, from a purely geometrical point of view. We want firstly to know if there is compatibility between the crystallographic lattices, and secondly to deduce an epitaxial strain tensor. We assume the existence of a coherent interface between the substrate and its oxide, i.e. we consider that the move from the crystallographic lattice of the substrate to the oxide's one is realized, from a crystallographic point of view, through a homogeneous transformation between the initial (substrate) and deformed (oxide) configurations. Two methods are used and compared:
- a first method giving, if it exists, the nature of the crystallographic compatibility at the interface (Hadamard condition [1], [2]),
- a second one giving a good coincidence criterion between the crystallographic phases (Bollmann's method) allowing to know the relative orientation of the oxide regarding the substrate, but also the induced strain value necessary to arrange the crystallographic lattices (epitaxy strain).

Furthermore, we compute using a FEM code the stress due to the connection between lattices of $UO_2$ and $U_3O_7$ and the diffusion of oxygen atoms in $UO_2$ using a coupled chemical / mechanical model. We will show that the thickness of $U_3O_7$ layer has an influence on the maximum stress inside this one, beyond which the layer cracks.

**Physical parameters.**

In the table below are given the parameters used in the different calculations and simulations presented in this paper.

|  | $D_1$ [cm$^3$/s] | $\eta^1_{ij}/\rho$ | T [°C] | E [GPa] | $\nu$ | Lattice parameters [Å] | | |
| --- | --- | --- | --- | --- | --- | --- | --- | --- |
|  |  |  |  |  |  | a | b | c |
| Ref. | [3] | [4] |  | [5] |  | [6,7] | | |
| $UO_2$ | $0.0055 \exp\left(\dfrac{-26.3}{RT}\right)$ | $-1.248 \cdot 10^{-5}$ | 300 | 200 | 0,32 | 5,47 (cubic) | | |
| $U_3O_7$ |  |  |  |  |  | 5,40 (tetragonal) | | 5,49 |

Table 1. Parameters of $UO_2$ and $U_3O_7$.

**Study of the epitaxy relations between a substrate $UO_2$ and its oxide $U_3O_7$**

**Hadamard condition.** In the case of a homogeneous transformation between two configurations (initial and deformed), we define, at any instant t, a vector function $\Phi$ written as an affine correspondence between two vectors $\vec{X}$ and $\vec{x}$ of a particle in the two configurations:

$$\vec{x} = \underline{F}(t)\vec{X} + \vec{c}(t). \qquad (1)$$

$\underline{F}(t)$ is the transformation gradient in $M_0$ given by $\underline{F}(t) = \nabla\Phi(\vec{X},t)$, with det($\underline{F}(t)$)>1. $\underline{F}(t)$ allows to link an elementary vector $\vec{X}$ describing $M_0$ in the initial configuration to its counterpart $\vec{x}$ in the deformed configuration. If we consider two domains $\Omega_1$ and $\Omega_2$ separated by a plan (normal $\vec{n}$) and undergoing continuous displacements, it must exist a condition on these ones, called compatibility condition at the interface, describing the fact that the two linear deformations inside each domain have a coherent interface whereas the gradients of these deformations are different both side of the interface (i.e. the two strains are different). In this way, let's consider the gradients ($\underline{F}_1(t)$ and $\underline{F}_2(t)$) of the homogeneous transformations in the two domains $\Omega_1$ and $\Omega_2$: these gradients are defined in each domain, using Eq. 1. The two corresponding equations have to be verified, at the same time, for any point of the interface (i.e. the two position vectors $\vec{x}_1$ and $\vec{x}_2$ are the same at the interface). These equalities lead to introduce a vector $\vec{a}$ corresponding to the difference between the transformations of vector $\vec{n}$ both side of the interface. We obtain then [2]:

$$\underline{F}_1(t) - \underline{F}_2(t) = \vec{a} \otimes \vec{n}. \qquad (2)$$

This equality corresponds to the compatibility condition at the interface (Hadamard condition). From a purely crystallographic point of view, for phases whose lattices are different, this condition allows to know if these phases are compatible, i.e. can coexist on each side of a coherent interface.

We applied this condition to the case of a $U_3O_7$ oxide layer with a (010) direction parallel to a (001) direction of a $UO_2$ substrate, perpendicular to the interface. We show that:
- a perfect interface between the substrate and the oxide can not exist a priori (there is no geometrical compatibility between the crystallographic lattices of $UO_2$ and $U_3O_7$).

- a perfect interface can exist between two $U_3O_7$ domains of different crystallographic orientations (possibility to obtain the coexistence of $U_3O_7$ domains): the principal axes of the lattices are then 90°-oriented relative to each other.
- A coherent but no perfect interface can exist between $UO_2$ and a mixture of two $U_3O_7$ domains of different crystallographic orientations separated by a perfect interface. The principal axes of the $U_3O_7$ lattices are oriented in the same direction than the ones of $UO_2$.

**Epitaxial strains: Bollmann's method.** This method [8] is used to determine epitaxial strains generated by the growth of an oxide on the surface of a polycrystalline strongly textured metallic substrate. Our work uses some theoretical results obtained by Salles-Desvignes [9] in our lab few years ago. This geometrical approach takes into account the position of lattice points. Let's consider a substrate "1" and its oxide "2" superimposed. To establish a criterion of the best fit between these two lattices, Bollmann interpenetrates their lattices and calculates, for each orientation of the oxide in relation to the substrate, the matrix of linear transformation $\underline{A}$. Mathematically, we can write:

$$\vec{X}_2 = \underline{A}.\vec{X}_1. \tag{3}$$

Where $\vec{X}_1$ and $\vec{X}_2$ represent the primitive unit cells of the substrate and its oxide, respectively. Then, matrix $\underline{A}$ must verify the complete coincidence of the two lattices. Bollmann then generates the "O" lattice, constituted by the lattice points verifying Eq. 3. In this case:

$$(\underline{I} - \underline{A}^{-1}).\vec{X}_0 = \vec{b}. \tag{4}$$

in which $\vec{X}_0$ is the representative matrix of the "O" lattice and b represents the translation coordinate system of the substrate lattice. So the matrix $\underline{A}$ is related to the primitive unit cell volumes $V_b$ and $V_O$ of the "b" and "O" lattices respectively. Indeed, Eq. 4 gives:

$$\det(\underline{I} - \underline{A}^{-1}) = V_b / V_0. \tag{5}$$

From the "O" lattice, it is possible to define a "O" cell, which contains points of the substrate and oxide lattices in nearest coincidence. So, the greater the volume of $V_O$, the more numerous the lattice points in nearest coincidence and consequently better is the fitting between the substrate and its oxide. As $V_b$ is constant, $V_O$ is in the inverse ratio to $\det(\underline{I}-\underline{A}^{-1})$. Then, a criterion of best fit between the two lattices can be given by: *the orientation of optimal coincidence between two lattices corresponds to the minimal determinant value of $(\underline{I}-\underline{A}^{-1})$, A being the matrix of linear transformation from one lattice to the other*.

Such a method is limited to very neighboured lattices (shape and size). In the case of distant lattices, Bonnet [10] generalized this method, replacing the "O" lattice by a "C" lattice, and the "b" lattice by a "B" lattice. These two new lattices can no longer be primitive. In the substrate and oxide lattices, let's consider two multiple lattice cells $M_1$ and $M_2$ in good coincidence. A new parameter has to be introduce: the density ratio $\Sigma_1$ (resp. $\Sigma_2$) corresponding to the multiplicity of $M_1$ (resp. $M_2$). So, the determinant of the matrix $(\underline{I}-\underline{A}^{-1})$ is equal to the ratio $V_{Mb}/V_{Mc}$, where $V_{Mb}$ and $V_{Mc}$ are the multiple lattice cell volume of "B" and "C" lattices. By comparison with Eq. 5, Bonnet writes:

$$\det(I-A^{-1}) = V_{Mb}/V_{Mc}. \tag{6}$$

By introducing $V_c$ (volume of the "C" cell), $V_1$ (primitive unit cell volume of the substrate such as $V_1=V_{M1}/\Sigma_1$) and $V_2$ (primitive unit cell volume of the oxide such as $V_2=V_{M2}/\Sigma_2$), one obtains:

$$V_c = \frac{V_1 V_2}{\det(\underline{I} - \underline{A}^{-1}).\det U_2}. \tag{7}$$

In this expression, $\det U_2 = V_{M2}$. Consequently, we can generalize the coincidence criterion by the following condition: "$\det(\underline{I} - \underline{A}^{-1}) \cdot \det U_2$ has to be minimal". For each orientation of the oxide in relation with the substrate, the computed transformation matrix A has to verify the criterion previously defined: the favored orientation of the oxide is the one that gives a minimal value of the criterion. A radar graph is then drawn and allows to determine the orientation of nearest coincidence. Furthermore, a value of the epitaxial strain tensor is determined using a classical Lagrange's definition for a macroscopic deformation.

In order to complete the results obtained previously and concerning the Hadamard condition, we applied first the criterion to the case of a $U_3O_7$ oxide layer with a (010) direction parallel to a (001) direction of a $UO_2$ substrate: the corresponding radar graph is given in Fig. 1a. We observe that the two lattices are in near coincidence if their principal axes are superimposed (0° between the two lattices with a symmetry of 90°).

In the case of a $U_3O_7$ oxide layer with a (100) direction parallel to a (111) direction of a $UO_2$ substrate, what was proposed by Allen for a $UO_2/U_3O_8$ system, the graph is given in Fig. 1b. The favored orientation of the oxide is then obtained for an $a_{U3O7}$ axis oriented at 40° to an $a_{UO2}$ axis.

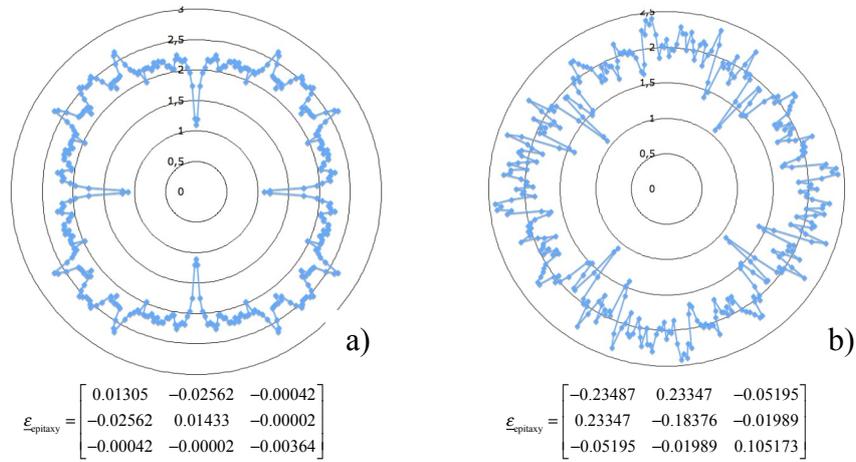

$$\varepsilon_{epitaxy} = \begin{bmatrix} 0.01305 & -0.02562 & -0.00042 \\ -0.02562 & 0.01433 & -0.00002 \\ -0.00042 & -0.00002 & -0.00364 \end{bmatrix} \qquad \varepsilon_{epitaxy} = \begin{bmatrix} -0.23487 & 0.23347 & -0.05195 \\ 0.23347 & -0.18376 & -0.01989 \\ -0.05195 & -0.01989 & 0.105173 \end{bmatrix}$$

Fig. 1. Radar graph of nearest coincidence criterion vs orientation of $U_3O_7$ in relation with $UO_2$: a) $(010)_{U3O7}$ parallel to $(100)_{UO2}$; b) $(100)_{U3O7}$ parallel to $(111)_{UO2}$.

**Modelling of a $U_3O_7$ layer on $UO_2$**

In order to highlight the chemical and mechanical coupling behaviour we will calculate the mechanical stress and the concentration of dissolved oxygen in $UO_2$ lattice. The background theory which give the influence of mechanical stress on the chemical diffusion was widely exposed in [9], [11], [12] and [13], and is based on previous works [14]. We recall here the main point used in our calculations. The diffusion coefficient depends on both the mechanical stress and the stress gradient:

$$D = D_0 \left[ 1 - \frac{M_0 \eta_{ij} c}{RT} \left( \frac{\sigma_{ij}}{c} + \frac{\partial \sigma_{ij}}{\partial c} \right) \right] \qquad (8)$$

where $M_0$ is the molar mass of the $UO_2$, $\eta_{ij}$ is the chemical expansion coefficient, $c$ is the oxygen concentration which is dissolved in $UO_2$ lattice, $\sigma$ is the mechanical stress, $T$ is the temperature and $R$ is the universal constant of gases.

We use here a strong coupling between concentration and mechanical stress. The stress locally modifies the diffusion coefficient and the oxygen concentration induces modifications in lattice parameters, which produces additional mechanical stress. Simulations will be made in two steps. The first step is to calculate the stress induced by the connection of $U_3O_7$ multi-domains layer and $UO_2$ substrate due to mismatch between their lattices parameters. The second step is to calculate the diffusion of oxygen in $UO_2$ in a coupled stress/diffusion model which takes into account the connection mechanical strain.

**The model**. According to the previous results, we consider a multi-domain of $U_3O_7$ (composed by a puzzle of simple domains) lying on a $UO_2$ substrate. The direction (001) in $UO_2$ is parallel to the direction (010) in $U_3O_7$, which corresponds to the case a) in Fig. 1. Thus, the $c(U_3O_7)$ direction lies in the plane of the interface. These domains have two different directions for the c-axis. We consider here that the first domain (called "type A") has it's a-axis parallel to the c-axis of the second domain (called "type B"). Fig. 2 shows such a puzzle of multi-domain. Boundaries between these domains are oriented 45° from the principal axes.

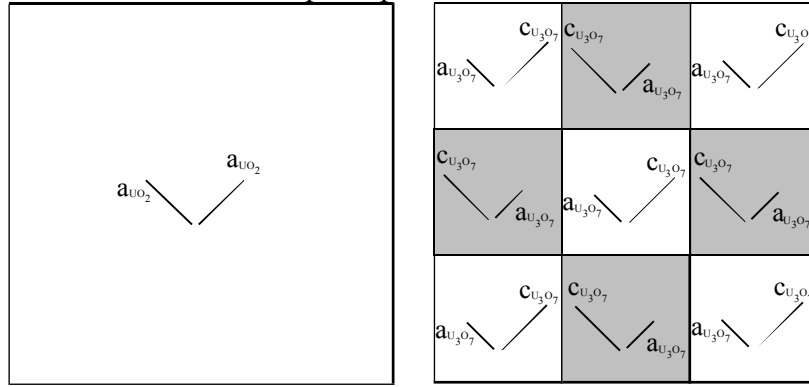

Fig. 2. $UO_2$ and $U_3O_7$ multi-domain sections: "type A" domains in white and "type B" domains in grey.

In our computations we use FEM code CAST3M. We simulate oxygen diffusion for several thicknesses of $U_3O_7$ upon infinite $UO_2$ substrate. *We do not simulate the growth of $U_3O_7$ layer, but we study influence of a given thickness on diffusion and stresses it induces, both in $UO_2$ and in $U_3O_7$ volumes.*

**Input parameters**. For $U_3O_7$ material, as it exists mainly in powder form, difficult to characterize, we assumed that the Young modulus and Poisson coefficient are the same as those of $UO_2$. We consider also that in $U_3O_7$ the concentration of oxygen is constant and the chemical expansion coefficient can be neglected. The initial conditions are zero oxygen concentration and non-zero mechanical stress (due to the connection between $U_3O_7$ multi-domain layer and $UO_2$ substrate). The chemical boundary conditions can be either concentration or flux imposed at the interface $U_3O_7/UO_2$. The mechanical boundary conditions consist on the elimination of rigid body displacements and on two symmetries to make calculations for only a quarter of the total volume.

**Results**. An example of connection stress due to the mismatch between lattices parameters of $UO_2$ and $U_3O_7$, lying on a section near the frontier $U_3O_7/UO_2$, is presented in Fig. 3.

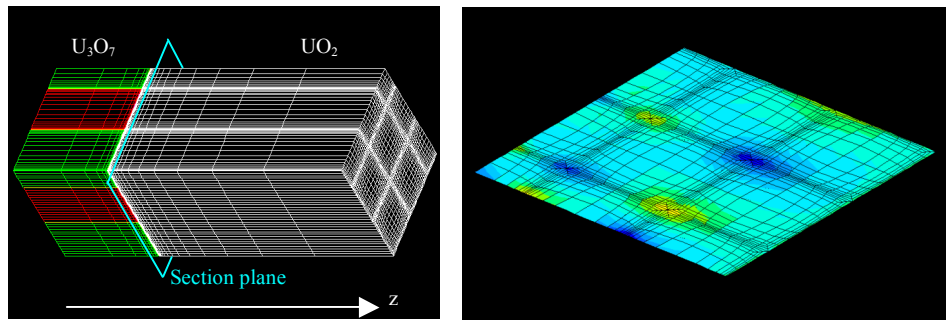

Fig. 3. FEM mesh (left) and $\sigma_{nn}$ connection stress (right) in a transversal section $UO_2$ side, at 1 nm from the boundary $U_3O_7/UO_2$; stress values are from -1 GPa to 1 GPa.

Calculations have been made for several thickness of $U_3O_7$ layer: the plot of variation of the $\sigma_{xx}$ stress is shown in Fig. 4. One can see that the maximal value of the stress, $U_3O_7$ side, is lower for a thickness of 50 nm and 100 nm than for a thickness of 200 nm. It seems to be a confirmation that stresses in $U_3O_7$ increase with the thickness of the layer. *We can say that a thick $U_3O_7$ layer induces maximal traction stress higher than a thin $U_3O_7$ layer. It can be an explanation of the experimental findings that the $U_3O_7$ layer cracks when its thickness reaches a critical value.*

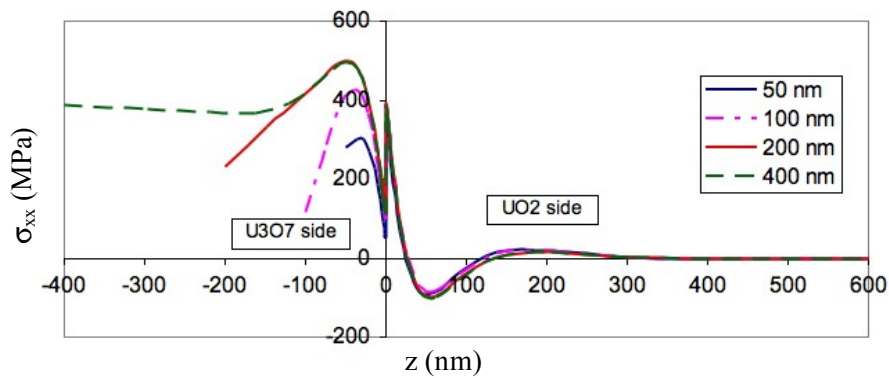

Fig. 4. $\sigma_{xx}$ stress for several thickness of $U_3O_7$ layer.

**Summary**

In this paper, a complete study of the influence of a $U_3O_7$ domain structure on cracking during the oxidation of $UO_2$ has been led. The possibility to have a multi-domain structure in the oxide $U_3O_7$, coexisting with the substrate $UO_2$, has been shown theoretically thanks to purely geometrical calculations on the crystallographic lattices of the substrate and its oxide (Hadamard relations and Bollmann's method). FEM simulations show, firstly, that the cracking can be due to the multi-domain structure rather than to the sample shape, and secondly, that a critical thickness of the $U_3O_7$ layer before cracking may exist.

**References**

[1] K. Bhattacharya, Acta Metall. Mater., Vol. 39 (1991), p. 2431.

[2] L. Hirsinger, N. Creton, C. Lexcellent, J. Phys. IV France, Vol. 115 (2004) pp. 111-120.

[3] W. Breitung, J. Nucl. Mater., Vol. 74 (1978) pp. 10-18.

[4] M. Dodé, B. Touzelin, Revue de Chimie Minérale, Vol. 9 (1972) pp. 139-152.

[5] V. Retel-Guicheret, F. Trivaudey, M.L. Boubakar, Ph. Thevenin, Nucl. Eng. And Design, Vol. 232 (2004) pp. 249-262.

[6] H.R. Hoekstra, A. Santoro, S. Siegel, J. Inorg. Nucl. Chem., Vol. 18 (1961) p. 166.

[7] Y. Saito, Nihon Kinzoku Gakkai-shi. Vol. 39 (1975) p. 760.

[8] W. Bollmann, Springer, Berlin (1970).

[9] I. Salles-Desvignes, T. Montesin, C. Valot, J.Favergeon, G. Bertrand, A. Vadon, Acta Mater., Vol. 38 (2000) pp. 1505-1515.

[10] R. Bonnet, Ph. D. Thesis, Université de Grenoble, France (1974).

[11] J. Favergeon, Ph. D. Thesis, Université de Bourgogne, France (2001).

[12] S. Garruchet, Ph. D. Thesis, Université de Bourgogne, France (2005).

[13] N. Creton, V. Optasanu, T. Montesin, S. Garruchet, L. Desgranges, Defect and Diffusion Forum, Vol. 289-292 (2009) pp. 447-454.

[14] F.C. Larché and J.W. Cahn, Acta Metallurgica, Vol. 30 (1982) pp. 1835-1845.